# A Bayesian social platform for inclusive and evidence-based decision making

Dr. S.K. Devitt, University of Queensland, corresponding author k.devitt@uq.edu.au

Ms. Tamara R. Pearce, Queensland University of Technology

Dr. Alok Chowdhury, Queensland University of Technology

Dr. Kerrie Mengersen, Queensland University of Technology

## Abstract

Against the backdrop of a social media reckoning, this paper seeks to demonstrate the potential of social tools to build virtuous behaviours online. We assume that human behaviour is flawed, the truth can be elusive, and as communities we must commit to mechanisms to encourage virtuous social digital behaviours. Societies that use social platforms should be inclusive, responsive to evidence, limit punitive actions and allow productive discord and respectful disagreement. Social media success, we argue, is in the hypothesis. Documents are valuable to the degree that they are evidence in service of or to challenge an idea for a purpose. We outline how a Bayesian social platform can facilitate virtuous behaviours to build evidence-based collective rationality. The chapter outlines the epistemic architecture of the platform's algorithms and user interface in conjunction with explicit community management to ensure psychological safety. The BetterBeliefs platform rewards users who demonstrate epistemically virtuous behaviours and exports evidence-based propositions for decision-making. A Bayesian social network can make virtuous ideas powerful.

# Contents

# Introduction

> …when it comes to the direction of human affairs, all these universities, all these nice refined people in their lovely gowns, all this visible body of human knowledge and wisdom, has far less influence upon the conduct of human affairs, than, let us say, an intractable newspaper proprietor, an unscrupulous group of financiers or the leader of a recalcitrant minority—H.G. Wells (1938)

In January 2021 a mob of supporters of Donald Trump stormed the Capital of the United States (Bergengruen and Time Photo Department, 2021). Despite no evidence of electoral fraud, and over failed 60 lawsuits to this effect, the rioters believed that their duty as Americans was to take back their country, to 'stop the steal' (Rutenberg et al., 2020, AP/Reuters, 2021). The mob believed that Joe Biden had been elected fraudulently, that democracy was at risk and that members of Congress had to be stopped certifying the electoral votes that would instate Joe Biden as the 46$^{th}$ president of the United States (McSwiney, 2021). False beliefs were incubated and amplified not by evidence, but by Donald Trump's posts on social media platforms, particularly Twitter and Facebook. Once posted on social media, Trump's messages went viral on social media and via a network of online forums and media creating a 'right wing echo chamber' (Tharoor, 2021).

There is no doubt that social media platforms sow disinformation and misinformation just as easily (perhaps much more easily) than true, verifiable information (Singer and Brooking, 2018). In the wake of the Capital riots, media commentors have reflected on issues of free speech and moderated content as they pertained to social media (Breton, 2021), wondering about the price society pays, particularly democratic societies, when lying becomes normalized (Tenove and McKay, 2021).

> The unrest in Washington is proof that a powerful yet unregulated digital space — reminiscent of the Wild West — has a profound impact on the very foundations of our modern democracies (Breton, 2021)

Where years of anguish and lament from ideologues has failed to change misinformation behaviours in the media and social media, corporate litigation has stepped in. Under the threat of defamation lawsuits, media outlets are now changing their behaviours (Brynbaum, 2021). Such lawsuits are having an immediate impact on misinformation narratives, e.g. during a rightwing media Newsmax interview 3$^{rd}$ Feb 2021, a host walked off camera to avoid engaging in discussions around unsubstantiated electoral fraud (MSNBC, 2021).

Against the backdrop of a social media reckoning, this paper seeks to demonstrate the potential of social tools to build virtuous behaviours online. If we believe that humans would benefit from incorporating philosophical theories into discourse and social knowledge structures, then social media platforms should be created, modified and updated based on our best normative theories in epistemology and the philosophy of science, rather than corporate monetisation metrics. That is to say, the impact of digital content on society should be proportional to the evidence we have for ideas and the comprehensiveness of this evidence. The more justified the ideas (e.g. climate change), the more these ideas should be promoted. If truth matters, then social media platforms must be neither contributor nor content neutral. And, if they are not neutral, then technology creators and their stakeholders must determine the manner and means of content management.

I take the following ingredients as important to creating good social platforms. First, we must accept humans for the sort of biased actors they are. Humans are myopic, overconfident and affected by contextual factors when they consider ideas (Montibeller and von Winterfeldt, 2015). Human behaviour is flawed—and that has to be ok. Second, the truth can be elusive and uncertain, perspectives subjective, evidence contradictory and opinions swayed by *ethos*, *pathos* as well as *logos* (Braet, 1992). Third, as communities, we must commit to values, and mechanisms that instantiate these values, to generate an overall society with greater epistemic virtues than our individual behaviours. Societies that use social platforms—either inputting and responding to data or using data produced on them—should value inclusivity, truth (and truth seeking); and should be receptive to evidence and evidence-based arguments. Platforms must also limit punitive actions and allow productive discord and respectful disagreement.

The ambition to create virtuous social information is not new. The history of information science is largely the instantiation of the dream to collect, collate, store, and access the world's best information and documents for the purposes of social good. From oral histories to the written word; shared taxonomies and indexing, to card catalogs and encyclopedias; databases, data mining, business intelligence and expert systems; and more recently recommenders, chat bots, and generative language models, humans have sought to store and share good information (Fivush and Haden, 2003, Krajewski, 2011, Wright, 2014, Dacome, 2004, Liao, 2003, Dale, 2021).

Hand-in-hand with virtuous information sharing, particularly since the invention of the printing press, is the parallel spread of propaganda and misinformation through pamphlets, books, newspapers and so forth (Burkhardt, 2017). The internet catapulted the potential to deceive and inform, leading scholars to interrogate the factors that need to be considered before one is justified in believing information online (Fallis, 2004, Bruce, 1997). Information literacy refers to the set of skills and epistemic framework that enable the identification of sources of information; how to access information, and then how to evaluate and use information effectively, efficiently, and ethically (Julien and Barker, 2009). Information literacy has renewed attention in light of the powerful impacts of 'fake news' since 2016 (Jones-Jang et al., 2021, Cooke, 2018). New normative frameworks to understand and proactively fight disinformation are emerging, e.g. Pamment and Lindwall (2021) provide normative debunking goals including:

> **Assert the truth:** Use established facts as a means of counteracting the negative impact of mis and disinformation
>
> **Catalog:** Develop a public record of falsehoods that are being spread by an actor in order to raise awareness of their behaviour and provide evidence of their actions
>
> **Expose:** Use identified mis- and disinformation as a starting point to expose the actors and networks behind the spread of false information

> **Attribute:** Collect evidence of an actor's behaviour in order to publicly shame them and support the imposition of costs to their actions
>
> **Build capacity:** Develop the skills and procedures to protect vulnerable institutions
>
> **Education:** Build societal resilience by educating the public about the tactics, techniques and procedures of disinformation

In this chapter we investigate whether mis- and disinformation can be fought using a social platform that resembles existing platforms, but simultaneously encourages virtuous information behaviours by its design.

The rise of social media in some ways has marked the demise of the document as a primary unit of information (Buckland, 1991, Wright, 2007). Rather than building up knowledge in expert systems, social media encourages ephemeral, unexpert ejaculations. Social media builds on human gossip mechanisms for shared belief, rather than co-constructing more faithful representations of reality. This chapter suggests new path for social media in an age of uncertainty and a hunger for evidence-based collective thinking. There is evidence that crowds can be wise, if the circumstances of deliberation and dissent are considered, and mechanisms of groupthink avoided (Solomon, 2006, Sunstein, 2011).

Social media success, we argue, is in the hypothesis. The document has long reigned as the unit of information with keywords, indexes and other signals indicated connections to other documents. In the platform we create, the primary unit of information is the hypothesis. Here documents are not intrinsically valuable, but valuable to the degree that they are evidence in service of or to challenge an idea for a purpose (Devitt, 2013). Such a reframing allows for and anticipates documents to be error-prone and variable in usefulness in accordance with the ambitions of Bayesian epistemology (Bovens and Hartmann, 2004, Hajek and Hartmann, 2009, Dunn, 2010, Gwin, 2011). Centering the hypothesis removes the barrier to using diverse information while limiting the influence of evidence used disproportionately or inappropriately. Traditional social media prioritizes the idea too, but to the detriment of evidence and expertise. Social media's infinite feed of assertions with little evidence creates almost the opposite information environment than that perfected by the book, the document, the card catalogue and the database.

The future of informed conversations requires far better utilisation of the global 'world brain'[1] of information through intuitive, yet structured social platforms. To this end a group of researchers have created a Bayesian social platform for evidence-based collective decision making which we articulate below.

# Social Media

The internet (more broadly), and social media (more specifically) has invited democratic participation in the espousement and evaluation of ideas. Wishing to remain impartial, social

---

[1] See H.G. Wells pre-internet, pre-Wikipedia vision of an updating encyclopedia in every library and institution in 'World Brain' (1938)

media companies have generally welcomed all who wish to register and share their data with them to monetise (Zuboff, 2019, Barnet and Bossio, 2020). Simple popularity metrics have been employed to adjudicate and share ideas, such as 'up voting' and 'starring' content; and retweeting and sharing content within or across platforms. But few features are built or deployed that explicitly work towards improving both the veracity or quality of information shared or the ability of users to effectively evaluate poor information or misinformation. Instead users share and like information amongst like-minded peers (Schmidt et al., 2017), reducing the friction of dissent and creating epistemic echo chambers. In-group messages expressing righteous or virtuous anger are propagated, while calm, moderate or evidence-based messages are shared less (Singer & Brooking, 2018). The science of human behaviours on current dominant social media suggests that, left to their own devices, humans are more likely to reinforce beliefs signalling social group membership/identification and less likely to collectively promote evidence-based beliefs.

This is despite a decade of empirical and theoretical social media research on ways people experience information on platforms such as Twitter and normative guidance for platform producers. For example, Zubiaga and Ji (2014) found that credibility perceptions of tweet authors played a significant role in how users trusted tweets. Basically, the more credible the 'tweeter', the more the tweet would be reshared.

Only after 14yrs has Twitter added a feature that asks users to employ metacognitive skills, to consider their actions, 'would you like to read the article before retweeting it?' In 2020, they ask this question if a user tries to retweet before reading a link (sharing based on trust), rather than opening the link (sharing based on knowledge)—see *Figure 1*.

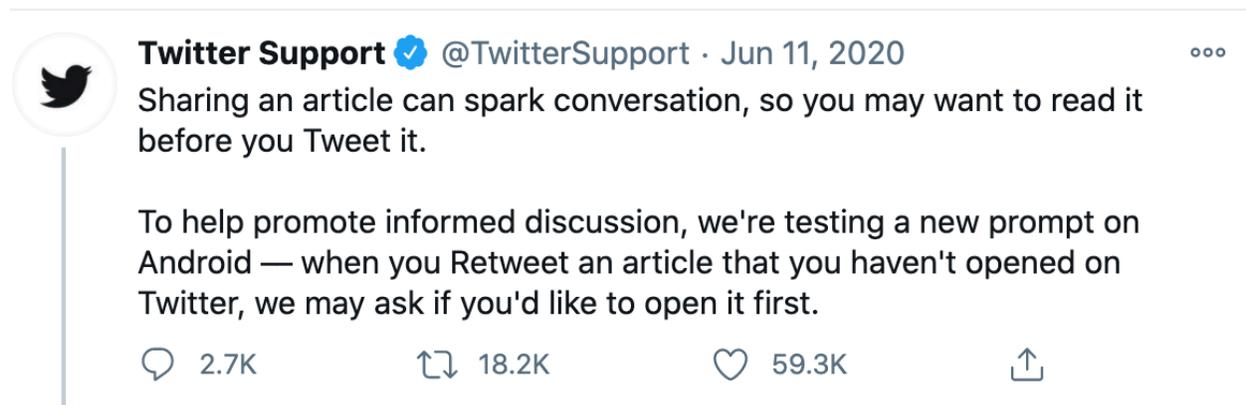

*Figure 1.* Twitter Support tweet explaining the new feature, a prompt to encourage informed discussion. See https://twitter.com/TwitterSupport/status/1270783537667551233?s=20

The experiment with some platforms (starting with Android) went extremely well, with users opening articles 40% more before sharing them, that Twitter has rolled out the feature across all platforms (Hatmaker, 2020). Twitter explains this feature because sharing an article can 'spark conversation' and opening articles (implied, 'reading articles') helps promote informed discussion—see *Figure 1*.

Devitt, S.K., Pearce, T.R., Chowdhury, A., Mengersen, K. (2021). A Bayesian social platform for inclusive and evidence-based decision making. [under peer review] M. Alfano, C. Klein and J de Ridder (Eds.). *Social Virtue Epistemology*. Routledge.

Social media has traditionally avoided censoring individuals (bad for business) and have allowed networks to grow and their advertising revenue to grow beside it, for example:

> At YouTube, we've always had policies that lay out what can and can't be posted. Our policies have no notion of political affiliation or party, and we enforce them consistently regardless of who the uploader is (Novacic, 2020).

Disregard for political affiliation has led to the rise not only of political extremism but has also made social media the locus of political action such as recruitment, propaganda and collective action. For example, Facebook, Twitter and YouTube were central in the rise of cyber jihadists and Isis (Awan, 2017). Facebook enabled warring militias in Libya's civil war to generate and sustain power (Singer and Brooking, 2018, Walsh and Suliman, 2018). While white supremacists and conspiracy groups such as QAnon in the United States have grown and strengthened with the comprehensiveness of open information on the internet and social media (Hannah, 2021). Social media companies do have guidelines to pull down content that includes hate speech, inappropriate content, support of terrorism, or spam. But, they also rely on inscrutable decision-making, large cohorts of preciously employed content moderators and automated tools (Ganesh and Bright, 2020, Roberts, 2019, Gillespie, 2018).

However, after the unprecedented mob attack on the US Congress 6 Jan 2021, incited by weeks of delegitimising the US election, Twitter first suspended the personal Twitter account of the President of the United States Donald Trump and then permanently deleted it when the user did not obey Twitter's governance rules. Facebook also deleted Trump's accounts and Apple and Google removed the social media app Parlour from its app stores. Amazon removed Parlour from its web hosting services. Within a week of the attacks thousands of accounts inciting violent insurrection against the US government were removed by Twitter and Facebook.

The question remains whether the solution to social media lies less in content moderation, and perhaps more in the way interaction occurs and information is used. Democratic participation needs to value inclusion and diversity, but also prioritise the knowledge and experience of experts and expertise. Evidence must be drawn from a defensible range of stakeholders and there must be a reasonable opportunity to submit ideas and evidence. Similar to the slow-food movement, future social media must gather and analyse data for propositions over longer temporal periods. The digital social epistemology movement must find a way to encourage interactivity, thoughtfulness and genuine engagement, while also mitigating human cognitive and affect limits, human biases and tendencies.

# Background

This chapter considers how groups of people might come together more effectively to understand a problem space and to propose actionable solutions.

## Social information processing

The field of social information processing has long questioned the role of social interaction on social information processing from in person office interactions to online virtual experiences (Festinger, 1954, Salancik and Pfeffer, 1978, Meyer, 1994, Ahuja and Galvin, 2003). Individuals are motivated to communicate with others in order to establish socially derived interpretations for events and their meanings when judgments are important, but evidence is ambiguous or non-existent and information complex (Salancik and Pfeffer, 1978, Meyer, 1994). Groups of people desire to fit in and will be motivated to agree with the group. With repetition, ideas are likely to convince individuals, that is make them believe them. Humans use social reasoning as a tool to make sense of uncertainty. Social platforms provide epistemic checking for groups. People will tend to believe what others in their group believe. If evidential reasoning is valued and social reasoning requires evidence, the group may collectively believe propositions for which there is corresponding evidence.

## Data-driven decisions

A problem that has arisen across social media and within traditional organisations is that while overt strategy might recommend 'data driven decisions' (Haller and Satell, 2020), in actual fact, decisions are largely made based on political will, trend, and biases arising from limited time and resources to evaluate ideas. Even when organisations use data for decisions, often data is incomplete, inaccurate, irrelevant or otherwise problematic to use to base decisions on (Provost and Fawcett, 2013). Data is rarely used by itself in raw form, but is transformed via human or machine interpretation, so when we speak of 'data' in this chapter, we mean data, models and algorithms; as well as whether data are classed as assertions (aka hypotheses) or evidence for or against hypotheses.

Whatever one defines data as, two things are true, 1) data is thought to be valuable and 2) data is difficult to use. What is a real method to use data to make decisions, even when it is partial, messy, of varying quality problematic? The method suggested in this chapter is highly pragmatic, yet grounded on solid philosophical foundations. The method allows a risk-based approach to data-driven decision-making, where stakeholders to the decision are 'at the table' and given a timeline to contribute to decisions, but that there is an end to deliberations and hand-wringing. There is also political heft to decisions proportionate to the diversity and range of stakeholders invited to contribute and the quantity and quality of contributions by said stakeholders. Unlike significance testing in the social sciences, there is no magic threshold of evidence under which truth can be presumed. But, following the tenants of Bayesian epistemology, beliefs ought to get stronger the greater the evidence there is to believe in them.

## Environments, letters, and online communities

Humans have been shaping their environments for hundreds and thousands of years to convey knowledge through acts such as path-making, cave painting, creating physical sequences for making or using tools or carving messages on objects or paper (Sterelny, 2012, Sterelny, 2003).

Devitt, S.K., Pearce, T.R., Chowdhury, A., Mengersen, K. (2021). A Bayesian social platform for inclusive and evidence-based decision making. [under peer review] M. Alfano, C. Klein and J de Ridder (Eds.). *Social Virtue Epistemology*. Routledge.

Letters formed the beginning of written dialogues between humans and are acknowledged as pivotal in shaping the beliefs of social groups (particularly dyads). Letter writing has affected the history of ideas such as Princess Elisabeth of Bohemia who wrote to Descartes 1643 to 1650 (Descartes, 1989). Email quickly took over the traditions of letter-writing in the 1990s and early 2000s, leaving digital rather than physical records of interactions. At the same time, online discussion boards created social communities to interact and share ideas. The rise of social media in the mid-2000s saw archival communications massively reduced. It remains incredibly (and intentionally) difficult for users to re-find the ideas they have expressed on platforms such as Facebook and Twitter. There has recently been a backlash of sorts against the ephemeral group interactions on Facebook and Twitter and a renewed interest in 1-on-1 engagement via apps such as Messenger. The role of social media in the future of communication remains 'up for grabs'. Still these written forms maintain a dialogue between people, a relationship, with no specific end date or event in mind.

## Brainstorms, workshops and conferences

Organisations and inter-organisational groups use the mechanisms of decision-oriented meetings (brainstorming sessions, strategy, or evaluative), workshops or conferences to build social epistemic communities. These interaction events are spatiotemporally limited to achieve particular outcomes. Since Covid-19, online meetings, workshops and conferences have become the standard for group interactions. But the tools used to experience these meetings often lack in-depth interactivity to mimic the experience of in-person events. Typical workshops and conferences encourage discussions between presentations, and it is often acknowledged that conversations, 'at the bar' is where and when intellectual progress really occurs.

There are assumptions made about the value of these meetings, sometimes explicit, though often implicit or taken for granted. There are two broad overlapping categories of supposed benefit from these meetings. One is social and about building human relationships through shared experience. The second is disseminating individual knowledge (testimony) and constructing group knowledge. Group knowledge can result in the production of co-authored publications including reports, journal papers and edited books, or single-authored publications that are more likely to refer to and cite the ideas of others invited to said workshops and conferences. There is a non-rigorous distinction between the workshop and the conference.

Workshops can be bespoke, idiosyncratic and useful for a point of time to achieve a specific end and those attending may never meet again. They tend to be more interactive, with greater emphasis on using tools such as sticky notes, white boards, mind-mapping software and design thinking to overcome the individual for the sake of collective production for a purpose.

Conferences tend to be reoccurring events to build an epistemic community over time. Attendees and presenters shape the future direction of the collective. The psychological pull of attending the same events year after year is to maintain social relationships and witness one's own part in shaping the direction of the group's thinking over time.

Devitt, S.K., Pearce, T.R., Chowdhury, A., Mengersen, K. (2021). A Bayesian social platform for inclusive and evidence-based decision making. [under peer review] M. Alfano, C. Klein and J de Ridder (Eds.). *Social Virtue Epistemology*. Routledge.

Now, more than ever, there is a gap in digital social tools that promote the epistemic aims of communities embarking in knowledge sharing and building. This is what has brought us to consideration of what justifies social platforms (or could justify social platforms)?

## Epistemic justification of social platforms

Ideas are posted to social platforms, but how is any information shared justified? Or to put it another way, what gives ideas and information authority, trustworthiness or credibility for decision-makers to progress decisions? Once we can identify what sorts of information we want to see on platforms, then we can consider how to advocate for virtuous online behaviours to manifest better information amongst participants and better management or treatment of this information buy decision-makers. This section will go through some of the main sources of justification for information pertinent to digital information sharing. Discussions we won't go into include those around internalism vs. externalism that seek to ground human beliefs against 'brain in a vat' style arguments.

For the sake of the chapter, we assume the following:

**P1.** **Realism:** basic human beliefs are, for the most part, grounded in perceptions and experiences in the external world that correspond with external reality, e.g. humans really see tables, chairs and trees (Devitt, 1997, Kornblith, 2002) and are not in skeptical conditions (Unger, 1978, Audi, 2010 Ch.13-14).

**P2.** **Digital Skepticism:** human beliefs are increasingly influenced by veristically-challenged online information environments that require skeptical vigilance (Cooke, 2018, Cooke, 2017). The saturation of AI-generated (Ippolito et al., 2020), false and misleading digital information increases minimally accurate, inaccurate and false beliefs depending on an agent's ability to curate, manage and correct information flows. Digital skepticism is particularly important information and behaviour promoted by media companies that seek to monetise user attention (Zuboff, 2019, Singer and Brooking, 2018) and information and behaviours suggested and reinforced by social peers (Eckles et al., 2016, Bailey et al., 2019) and echo chamber effects (Quattrociocchi, 2017, Cinelli et al., 2020a, Cinelli et al., 2020b).

**P3.** **Justification**: beliefs ought to have both a justified foundation (e.g. via perception, memory, expert testimony) and ought to cohere with other well-justified beliefs (Goldberg, 2012, BonJour, 2017). Information found in books and online need to be verified and justified on a case-by-case basis, but influenced by features such as authority, plausibility and support, independent corroboration, and presentation (Fallis, 2004, Fallis, 2008, Fallis, 2006, Zubiaga and Ji, 2014).

**P4.** **Social Epistemology:** ought to recommend error-correction mechanisms including overriding:
   a. singular inaccurate beliefs or poorly grounded beliefs of an individual, e.g. where an individual asserts a proposition for which they lack sufficient evidence,

> b. systematic inaccurate or poorly grounded beliefs, e.g. an individual or a group of individuals reliably assert propositions with misaligned correspondence with reality or for which they lack sufficient evidence

Combining these premises, we form a conception of humans interacting in information environments where their connection to reality via traditional modes such as visual perception and memory are grounded by virtue of being evolved to live and succeed in the real world (P1). Yet, human beliefs are increasingly under threat from the deliberate or incidental misinformation from online information environments (P2). In order to be justified in their information habits, humans must develop justified methods to find, sort and evaluate information sourced from a variety of sources (P3). The endeavour to improve epistemic habits is best done within physical and digital social groups (P4). The ambition then is to create digital infrastructure that provides the sort of justification that holds up to the highest epistemic standards. The benefit of digital tools is that time can be spend honing them over time against our best normative theories.

# An evidence-based social platform

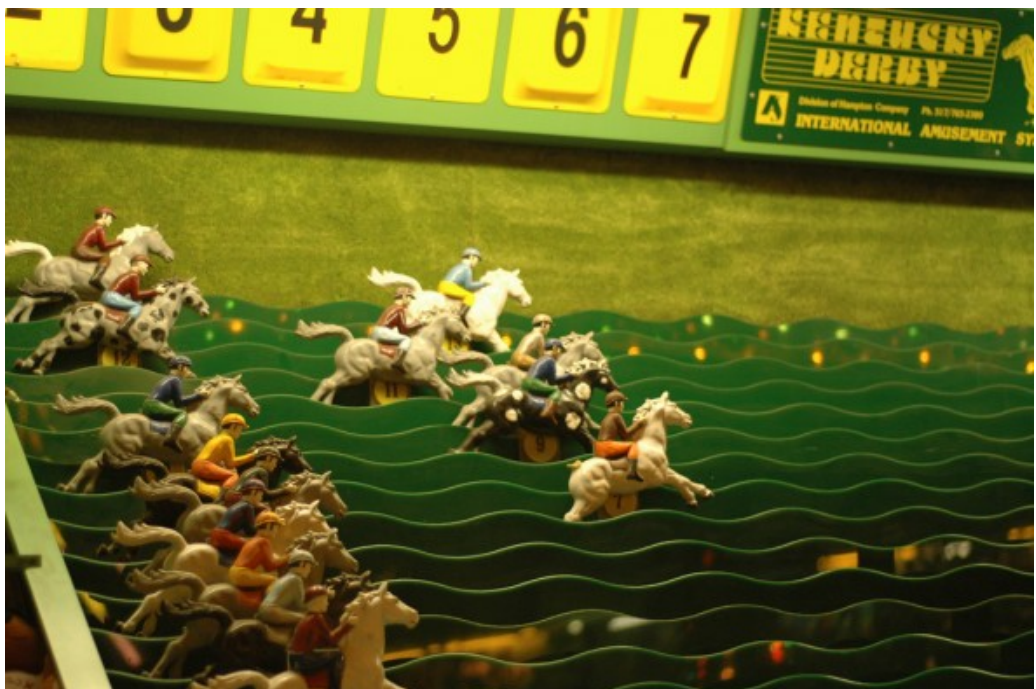

Figure 2: 20th Century horse race. Image: Casey Hibbard (25 March 2010)
https://www.compelling-cases.com/how-case-studies-get-done-one-leg-at-a-time/

Researchers at Queensland University of Technology (authors of this chapter) set out to make an evidence-based social platform that builds virtuous social information behaviours using interaction mechanisms that instantiate epistemic norms (Devitt et al., 2018). The researchers have diverse backgrounds from philosophy and cognitive science; business innovation and design; Bayesian statistics; machine learning and information technology. By encouraging social and evidence-based behaviours, the platform sought to build more scientific and inclusive digital cultures. Beginning as a research project, the team were funded by industry and grants to develop a minimal viable product (mvp) and then minimal marketable product (mmp) for market, creating a start-up around the platform 'BetterBeliefs'[2].

At its core, BetterBeliefs imagines ideas as hypotheses, representing by horses competing in a 'hypothesis horse race'. In order to progress in the race, the horses are fuelled by evidence, a little bit like the 20$^{th}$ C. carnival racing game where metal horses compete based on the number of interactions they receive from players (see *Figure 2*.). We thought it would be a breakthrough if data was connected to and presented for or against hypotheses, and data was psychologically engaging, rather than stored in databases hoping for a query to dig it up.

The core functions of the platform for users are:

- Submit hypotheses for consideration
- Submit evidence for and against hypotheses
- Vote on hypotheses to signify approval or disapproval
- Rank the quality of evidence provided for and against hypotheses
- Make a decision based on the degree of belief and weight of evidence of a hypothesis

## The business case

Organisations ineffectively use the data sets available to them and fail to maximise the value of expensive business intelligence systems (Drucker, 1999, Sharma and Djiaw, 2011, Richards et al., 2019). While organisations use business intelligence well for budgeting, financial and management reporting, they don't use them for corporate level decision-making (Richards et al., 2019).

As an Academic start-up dependent on industry-funding, the team needed the platform 'to sell', to have a clear value-proposition for business. We found evidence that social decision-making and innovation was good for business. For example, crowdsourcing using information systems can support management decision making through several stages of solving a problem (Chiu et al., 2014, Ghezzi et al., 2018, Lindič et al., 2011) such as:

1. **Intelligence** (e.g. search, prediction and knowledge accumulation),
2. **Design** (e.g. idea generation and co-creation) and
3. **Choice** (e.g. voting and idea evaluation) which lead to implementation.

However crowdsourcing can be a double-edged sword, particularly regarding problematic issues such as crowd attitudes and motives; and groupthink and other human biases (Chiu et al., 2014). Crowdsourcing using social platforms may help mitigate some biases in decision making for innovation but may introduce or exacerbate other biases depending on both platform features and how the platform is used (Bonabeau, 2009).

---

[2] http://betterbeliefs.com.au

Devitt, S.K., Pearce, T.R., Chowdhury, A., Mengersen, K. (2021). A Bayesian social platform for inclusive and evidence-based decision making. [under peer review] M. Alfano, C. Klein and J de Ridder (Eds.). *Social Virtue Epistemology*. Routledge.

Enterprise Social Media (ESM) is another information system that is a potential mechanism to share ideas across organisational silos, connect people and ideas, and enable innovation. Although the context of ESM is vastly different to commercial social media platforms discussed earlier, the literature on ESM shows that some of the decision making risk factors for social platforms translate across domains with echo chamber effects and biases including balkanisation and groupthink being highlighted as issues (Leonardi et al., 2013, Leonardi, 2014).

The business innovation literature revealed that high ideation rates (having lots of ideas) correlate with growth and net income across organisations. More specifically, there were four key elements essential to high ideation rates (Minor et al., 2017):

- **Scale** (more participants)
- **Frequency** (more ideas)
- **Engagement** (more people evaluating ideas), and
- **Diversity** (more kinds of people contributing

Designing a platform that encouraged these elements of ideation is a social platform that also addressed the thorny issue of effective, evidence-based decision making for innovation led to the creation of BetterBeliefs.

## How does it work?

To design BetterBeliefs, rather than reinvent the wheel of interaction, we selected intuitive mechanisms from existing social media and peer evaluation (e.g. Facebook, Twitter and Reddit). The essential functions of social media are:

1. Adding ideas (e.g. text, photos)
2. Responding to the ideas of others (e.g. indicating approval by upvoting or clicking on an icon or emoji, replying in a comments section

On our platform you can 'add new hypothesis' (see Figures 3 & 4) just like you can create a new tweet on Twitter. But, we built in new significance to the 'post' and 'like' functions to motivate individuals to think scientifically about claims relevant to their group.

A well-formed hypothesis is a simple proposition that a reasonable person can either agree or disagree with.

> E.g. Dogs ought to be the only companion animal allowed on domestic flights inside an aeroplane cabin

When forming hypotheses, we encouraged users to use words that imply what is obligatory, permissible, or forbidden, such as:

> Only, most, all, some, many, never, ought, permitted, should, can, needs, should not, cannot, may be, occasionally, sometimes, ought not, in some cases

Users add a hypothesis by giving it a title, a tag, some detail and then adding supporting or refuting evidence.

Devitt, S.K., Pearce, T.R., Chowdhury, A., Mengersen, K. (2021). A Bayesian social platform for inclusive and evidence-based decision making. [under peer review] M. Alfano, C. Klein and J de Ridder (Eds.). *Social Virtue Epistemology*. Routledge.

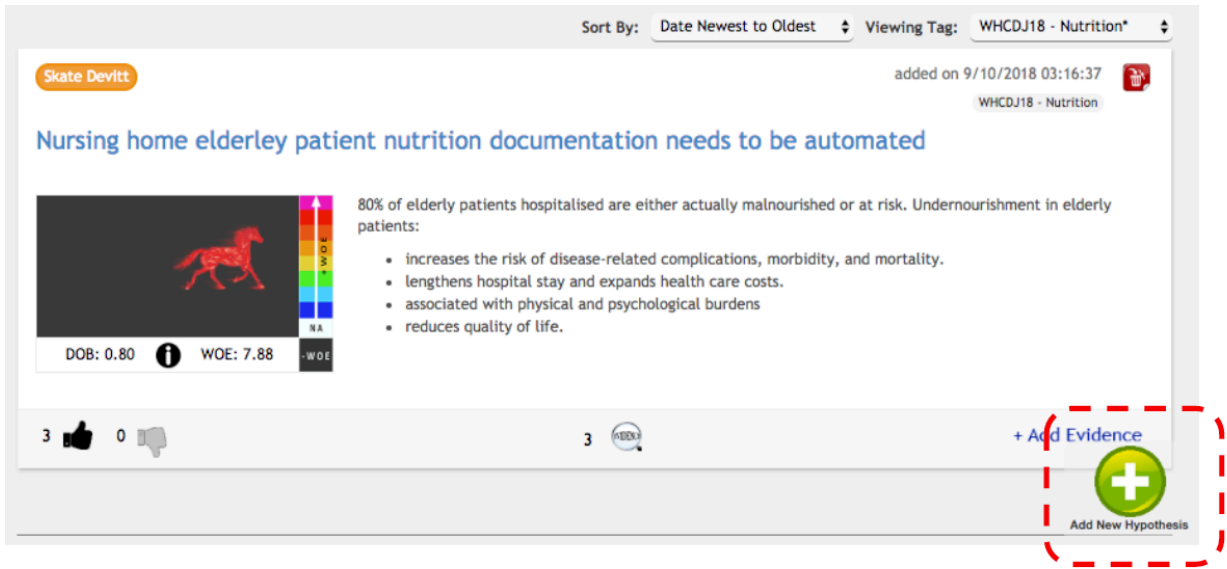

Add your ideas to the platform

Figure 3 Add a hypothesis to the BetterBeliefs platform

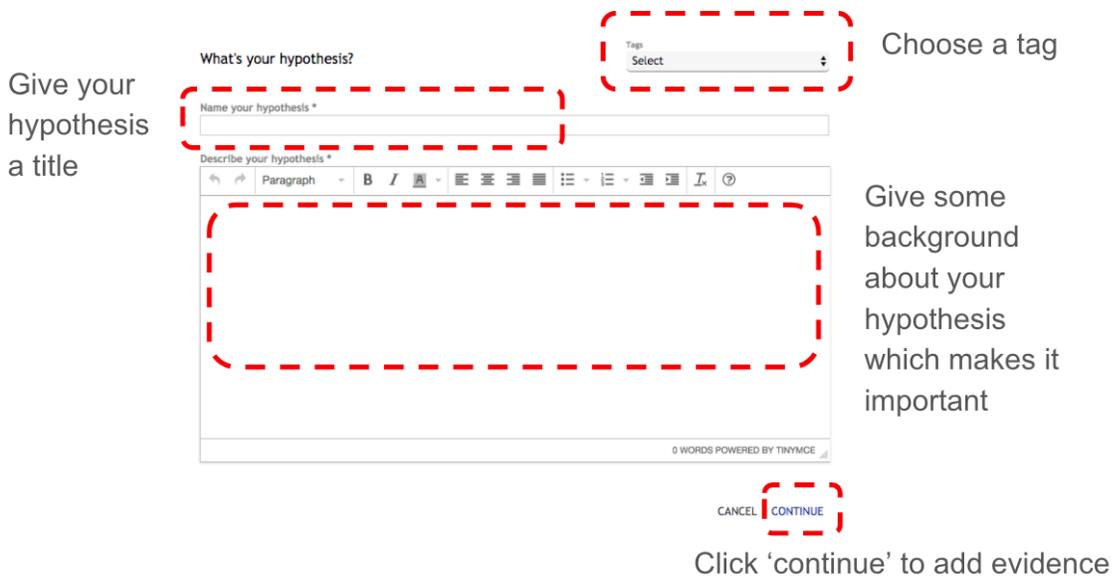

Choose a tag

Give your hypothesis a title

Give some background about your hypothesis which makes it important

Click 'continue' to add evidence

Figure 4 Detail your hypothesis

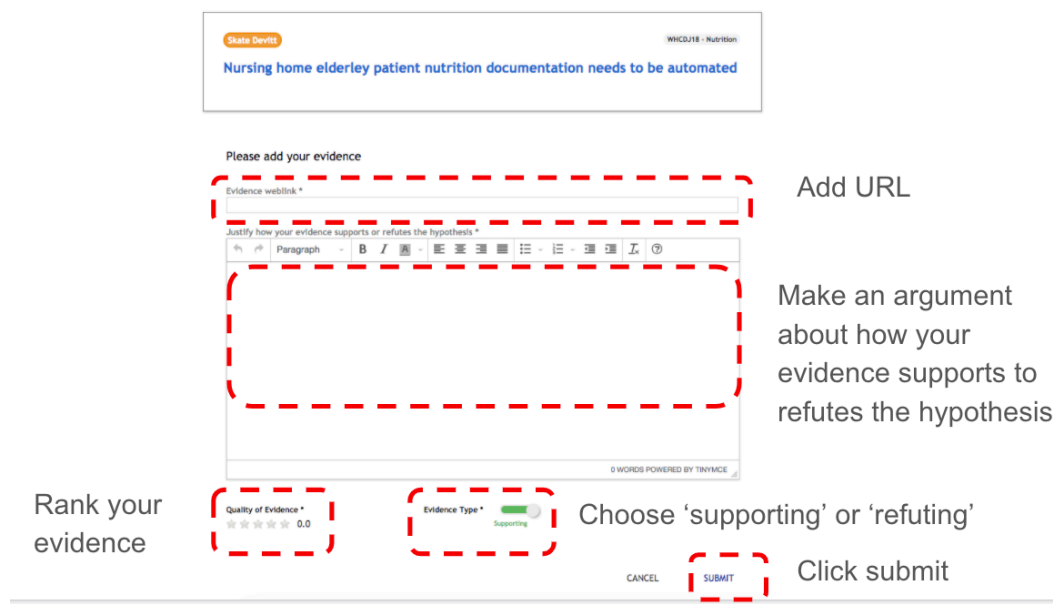

Figure 5. Add supporting or refuting evidence

When users add evidence (see Figure 5), they provide a URL, a brief argument that explains how their evidence supports or refutes the hypothesis (e.g. by example, abduction, analogy, defeasible, induction, deduction), rank their evidence and identify whether their evidence supports or refutes the hypothesis.

Note encouraging refuting evidence is a key part of BetterBeliefs that we believe no other social platform offers as a mechanism for epistemic evaluation.

Once the platform has hypotheses and evidence, the 'newsfeed' view shows users flaming horses and offers an opportunity to 'thumbs up' or 'thumbs down' the horses.

The degree of belief in the horse is represented by the position of the horse in the black 'racing box'. A horse to the left-hand side is poorly believed in. A horse to the right-hand side is 'winning the race', aka is highly believed in. However, just because a horse is on the right-hand side is not sufficient for a win—they need evidence too.

To that end, the horses change colour depending on the weight of evidence for or against them. White horses lack sufficient evidence. Pink horses have much evidence. Blue horses lack evidence. Black horses have evidence largely against them.

For example, a pink horse galloping to the right-hand side of the black box would be a good pick for decision-makers to progress. Whereas a white horse is better ignored until more interactions have occurred on it. In fact, a hypothesis will not turn from white to coloured until multiple users have interacted on the hypothesis in terms of both evidence and voting it up or down—See Figure 6.

Devitt, S.K., Pearce, T.R., Chowdhury, A., Mengersen, K. (2021). A Bayesian social platform for inclusive and evidence-based decision making. [under peer review] M. Alfano, C. Klein and J de Ridder (Eds.). *Social Virtue Epistemology*. Routledge.

Figure 6 vote hypotheses up or down

This simple interaction is the basis of the 'degree of belief' metric. In aggregate, an organisation or group can understand how much belief there is in a proposition—see Figures 7, 8 & 11.

Devitt, S.K., Pearce, T.R., Chowdhury, A., Mengersen, K. (2021). A Bayesian social platform for inclusive and evidence-based decision making. [under peer review] M. Alfano, C. Klein and J de Ridder (Eds.). *Social Virtue Epistemology*. Routledge.

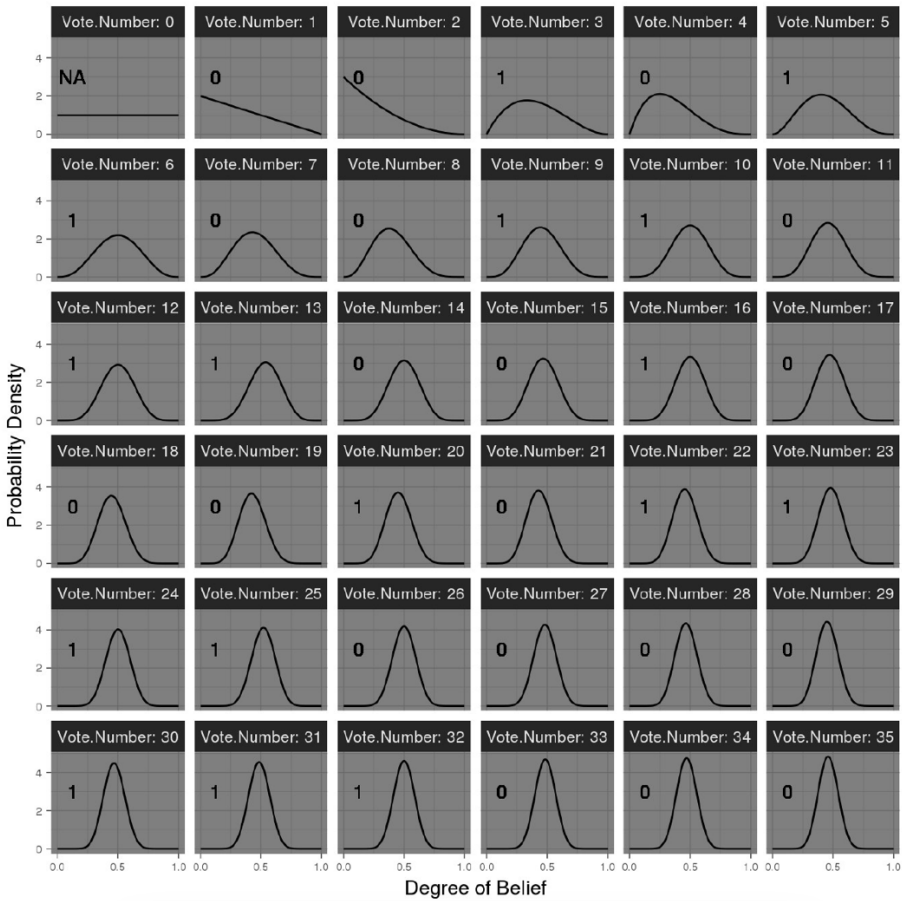

Figure 7. How the probability distribution changes for degree of belief over the first 35 votes on a hypothesis. Credit: Dr Benjamin R. Fitzpatrick

The degree of belief (DoB) metric takes the total upvotes and downvotes to create a likelihood that a hypothesis is true given user belief in it using Bernoulli-Beta distributions with 95% credible intervals represented to users. Our confidence in the degree of belief score increases the more users vote hypotheses 'up' or 'down (see Figures 7 & 8)

Devitt, S.K., Pearce, T.R., Chowdhury, A., Mengersen, K. (2021). A Bayesian social platform for inclusive and evidence-based decision making. [under peer review] M. Alfano, C. Klein and J de Ridder (Eds.). *Social Virtue Epistemology*. Routledge.

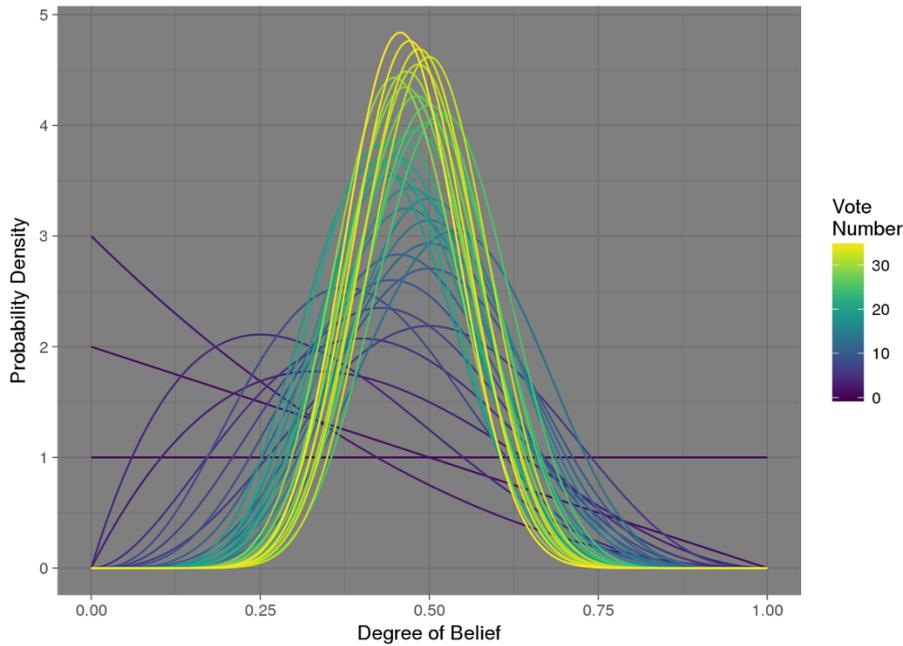

Figure 8. The change to probability density as votes are made on hypotheses. Credit: Dr Benjamin R. Fitzpatrick

The sum of evidence (supporting and refuting) plus the quality of evidence added forms the basis of the 'weight of evidence' score—see Figure 9.

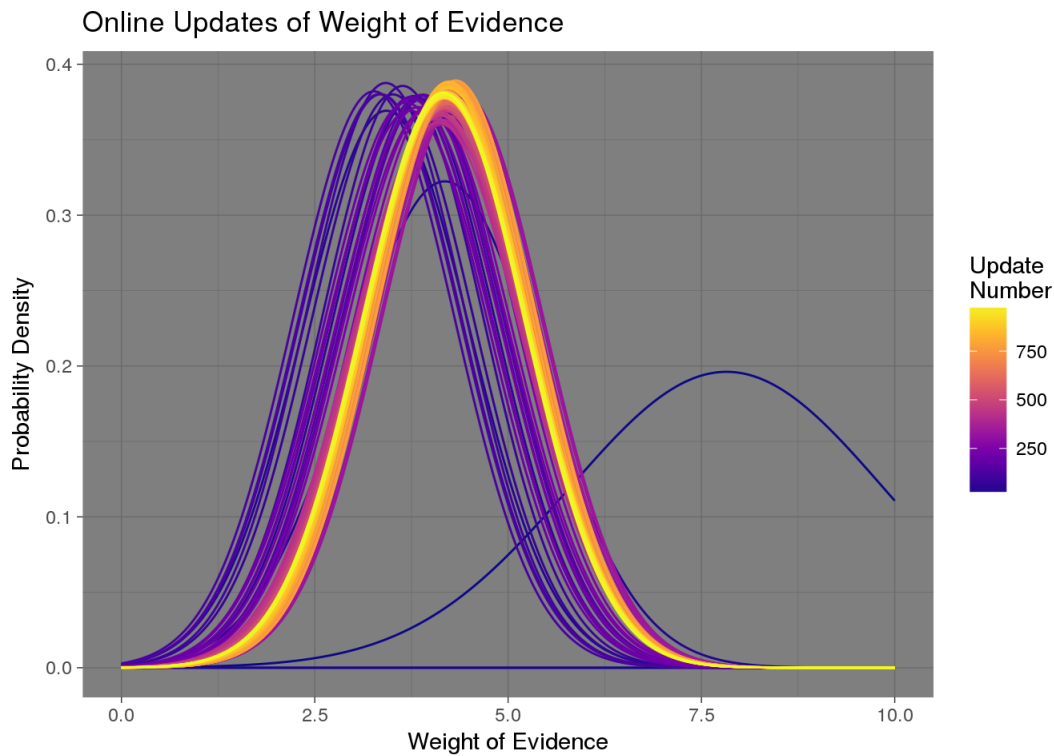

Devitt, S.K., Pearce, T.R., Chowdhury, A., Mengersen, K. (2021). A Bayesian social platform for inclusive and evidence-based decision making. [under peer review] M. Alfano, C. Klein and J de Ridder (Eds.). *Social Virtue Epistemology*. Routledge.

Figure 9. The probability density of the weight of evidence as number of supporting and refuting evidence items of varying quality increases. Credit: Dr Benjamin R. Fitzpatrick.

Not all evidence is created equally, so the quality of each piece of evidence must be evaluated to the degree that it supports or refutes hypotheses. When designing the platform, the researchers benefitted from work in statistical science as well as information science on the qualities of information that make it valuable (see *Table 1.*). The statistical methods that underpin the platform are currently not available to the public.

*Table 1* Dimensions of information quality and contributing factors for each dimension (Arazy and Kopak, 2011, Mai, 2013)

| **Dimension of Information Quality** | **Contributing Factors for Each Dimension** |
|---|---|
| *Credible* | Authentic, Believable, Reliable, Trustworthy, Authoritative |
| *Accurate* | Correct, True, Valid |
| *Relevant* | Contextual, Appropriate |
| *Comprehensive* | Complete, Objective, Neutral, Balanced |
| *Recent* | Current, Up-to-date |
| *Informative* | Understandable, Useful, Usable, Good |

They derived six dimensions: credible, accurate, relevant, comprehensive, recent and informative. During the initial design phase, the team considered inviting users to rate evidence on *each* dimension, but quickly felt that this would prove too taxing, generating an unwieldy user experience. In the end we created on a single star ranking (see *Table 2.*) that allowed users to rank evidence based on any combination of dimensions they felt was relevant to the rank. The team felt that the quality of evidence ranking, in aggregate, would produce 'better beliefs' for the collective than not having the ranking or requiring too much effort.

Devitt, S.K., Pearce, T.R., Chowdhury, A., Mengersen, K. (2021). A Bayesian social platform for inclusive and evidence-based decision making. [under peer review] M. Alfano, C. Klein and J de Ridder (Eds.). *Social Virtue Epistemology*. Routledge.

*Table 2.* Guide to ranking evidence items on BetterBeliefs

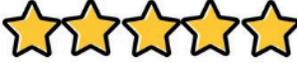

Once users have interacted with both hypotheses and evidence items the Evidence Engine produces the degree of belief (DoB) and weight of evidence (WoE) metrics—see figures 10 and 11 and Table 3.

The degree of belief is represented between 0.0-1.0, where 1.0 indicates 100% belief, absolute certainty in a hypothesis 0.5 indicates genuine uncertainty and 0.0 indicates absolute disbelief.

The weight of evidence is on a linear scale with no upper end limit. This choice is because theoretically there can always be further items of evidence that might increase the likelihood that a hypothesis is true. In reality, users engage with the platform for a finite period of time and there is a limit to the quality and quantity of evidence available to decision makers. Users can view the outputs of the Evidence Engine through the 'decision dashboard'—see Figure 11.

Devitt, S.K., Pearce, T.R., Chowdhury, A., Mengersen, K. (2021). A Bayesian social platform for inclusive and evidence-based decision making. [under peer review] M. Alfano, C. Klein and J de Ridder (Eds.). *Social Virtue Epistemology*. Routledge.

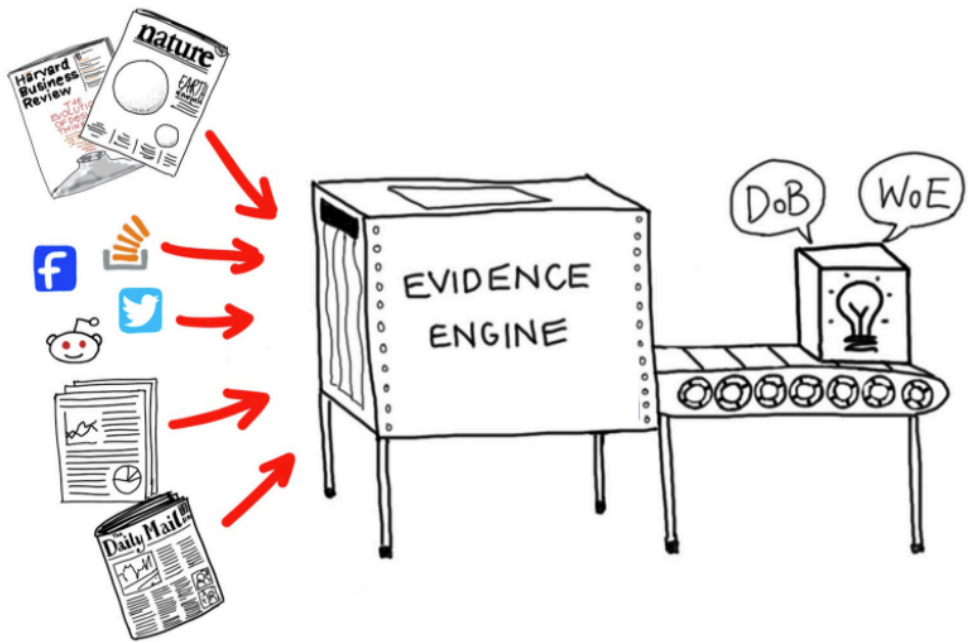

*Figure 10.* Users add many kinds of evidence to the platform to support or refute hypotheses. This information is translated into degree of belief and weight of evidence metrics

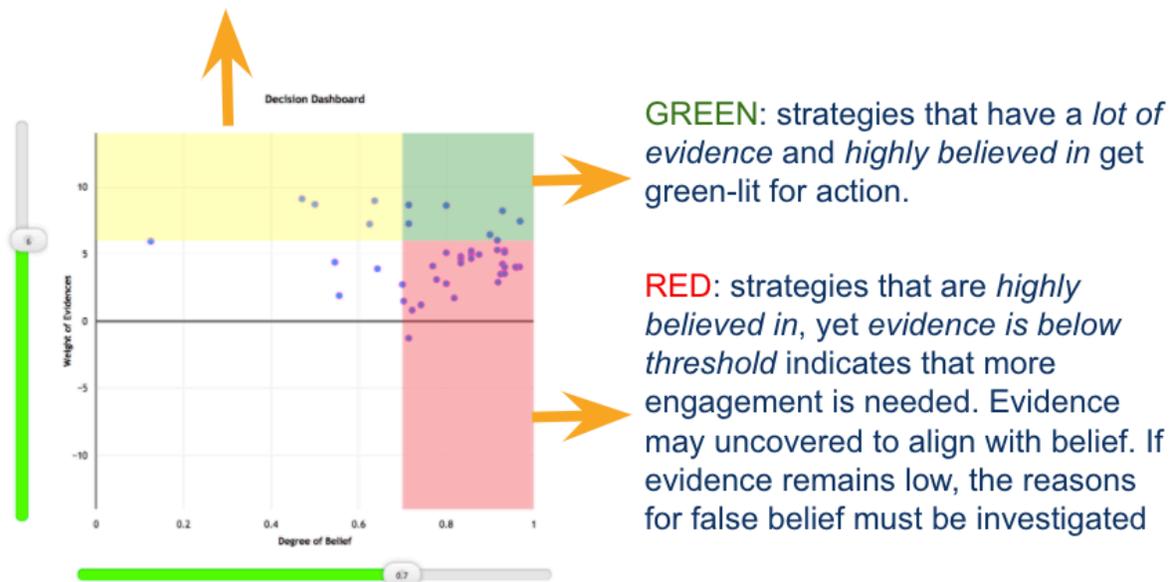

Figure 11. The Decision dashboard represents weight of evidence along the y-axis and degree of belief along the x-axis. Hypotheses are sorted into groups: green, yellow, red and white.

Devitt, S.K., Pearce, T.R., Chowdhury, A., Mengersen, K. (2021). A Bayesian social platform for inclusive and evidence-based decision making. [under peer review] M. Alfano, C. Klein and J de Ridder (Eds.). *Social Virtue Epistemology*. Routledge.

Table 3 Breakdown of decision quadrants: green, red, amber and white

| CATEGORY | DESCRIPTION |
|---|---|
| **GREEN** | The green box represents hypotheses that are 'greenlit for action' because they meet the decision-makers' threshold for both evidence and belief. Note that the decision maker can use the sliders to change the threshold depending on their own view of what is important for their decision and the consequences for making the decision. If it is a low-risk decision and/or a cheap or easy consequence from the decision, then the decision-maker may set a low threshold. However, if a decision has a lot of risk or the consequences of the decision may involve great costs or time, then the decision maker may require a higher threshold. In each case, due to the inevitable incompleteness of the evidence and limitations of contributors, decision makers will need to satisfice their choice—do 'enough' under limitations rather than optimise. They may make threshold decisions based on the number of hypotheses that end up in the green box and/or change the parameters of actions once the decision is made, e.g. if all hypotheses are insufficiently evidenced under one reward program, then instead of offering, say seed grants to highly believed hypotheses, they offer a 'revise-and-resubmit' to those landing in the green box. |
| **RED** | The red box represents hypotheses that are highly believed in yet lack sufficient evidence. A red hypothesis gives the pulse of belief and emotional buy-in. Red hypotheses mean different things depending on the expertise and diversity of participants. If participant intuitions are based on experience, decision makers might divert funds or resources to interrogate why hypotheses are highly believed yet short of evidence. It might be that evidence exists to back up high degrees of belief but are have not been added the platform. Or it might be that beliefs are in fact not sufficiently justified and there is only supposition. Either way decision makers can request users to seek out better evidence for their beliefs or suggest that they downgrade their degree of belief to be commensurate with their evidence. |
| **AMBER** | The amber box represents hypotheses that are have ample evidence, but are not highly believed in. An organisation may wish to<br>1. conduct information or education campaigns to communicate evidence in favour of these beliefs.<br>2. engage in safe social discussions to combat cognitive dissonance—where individuals are aware of evidence against their beliefs, but struggle to change them (Beck, 2017).<br>3. encourage unbelievers to add counterevidence to the platform to better justify their beliefs. |
| **WHITE** | The white box represents hypotheses that are contentious: have mixed belief or low belief and/or have mixed or limited evidence. There is a diversity of responses to these hypotheses, but the decision-maker is unlikely to progress actions on the basis of incomplete or contentious hypotheses. Still the controversy itself is evidence for decision-makers (Christensen, 2009). True disagreement offers an opportunity to rethink, reframe and reinvest in seeking good reasons for ideas and taking seriously arguments against them. |

Finally, users of BetterBeliefs can search the platform for keywords, they can filter hypotheses by recency, degree of belief, the number of evidence items and weight of evidence. Analytics are also available for each hypothesis to get a view of a hypothesis over time.

# Design principles

The platform is designed to:

1. Motivate the creation of more relevant options (hypotheses)
2. Evaluate options by explicitly linking to evidence
3. Harness stakeholder justifications for how evidence supports or opposes these hypotheses
4. Rank evidence to the degree it is a) quality, b) relevant to hypothesis it's connected with, and c) informative to evaluating hypotheses it is linked with.
5. Inform decision-makers about stakeholder ideas and vice versa
6. Harness the attraction of social media to teach the scientific method
7. Empower groups to make strategic decisions based on stakeholder generated and evaluated hypotheses

The platform aims to

- Help strategic decision making
- Make assumptions explicit
- Map social beliefs, knowledge
- Ignite collective expertise of stakeholders
- Generate new hypotheses for innovation
- Systematise rational skepticism using social Bayesian epistemology and statistics

## Evidence

A central justification for having beliefs is the degree of evidence one has for them. Much of the history of thinking about evidence in epistemology is about individual rather than collective beliefs. For example, if a person sees 40% chance of rain on the weather report, they should have some degree of belief that it will rain today. The more evidence a person has the more they should believe a proposition. The greater the risk of having a belief, the more evidence a person should have for that belief. A person should firmly believe a proposition when they have sufficient evidence for it. For example, Jill definitely believes that it is raining when she feels rain falling on her shoulders. In general, awareness of one's evidence for beliefs is considered a good thing, but the degree to which reflective access is required to be justified in believing is debated (see Dougherty, 2011). A reliablist may believe that dog know that it is raining even if the dog does not understand how she came to this belief (perhaps it was the smell of petrichor and the sound on the roof). It's a premise in this paper that justification in social epistemology stems from reliably formed beliefs, rather than depending on contributors to have reflective knowledge of what justifies their beliefs. That being said, the social platform discussed in this paper has only been used in use cases where the invitees were carefully selected for expertise. The platform tries to optimise the likely results with a bias towards diversity and expertise, plus the requirement that all ideas added to the platform are evidence-based. The platform also empowers a decision-maker to adjust the threshold of both the degree of evidence and degree of belief required for a hypothesis to be selected for some future action.

## Beliefs

Traditional epistemology tends to treat beliefs as 'all-or-none', either a person believes in p or ~p. Beliefs in a functionalist theory of the mind play a certain functional role in the cognitive architecture of an agent. If the agent believes p, then they act as though p were true. Beliefs provide scaffolding to guide and constrain behaviours as well as generating other cognitions such as desires or hopes. For example, if a person believes the US election was fraudulent, then they may storm the capital to take back democracy. Beliefs drive behaviour, even if they are objectively false. Bayesian epistemology takes a different perspective on beliefs. Instead of all-or-none, typical beliefs exist (and are performed) in degrees, rather than absolutes, represented as credence functions. This idea stems from Thomas Bayes who argued that our success in the world depends on how well credence functions, represented in our minds, match the statistical likelihoods in the world (Bovens and Hartmann, 2004). This statistical approach to beliefs enables agents to hold multiple beliefs, even contradictory beliefs in their minds at the same time with less certainty. There is evidence that the mind is Bayesian to a certain extent, using adaptive inference to change credence functions in response to evidence (Clark, 2015, Perfors, 2012, Gopnik and Wellman, 2012).

## Reducing biases

We aim to reduce cognitive and motivational biases (Kahneman, 2011, Montibeller and von Winterfeldt, 2015) by:

- Providing multiple and counter anchors
- Prompting employees to consider reasons in conflict with anchors
- Building explicit probability competence
- Providing counterexamples and statistics
- Capitalising on multiple experts with different points of view about hypotheses
- Challenging probability assessments with counterfactuals
- Probing evidence for alternative hypotheses
- Encouraging decision makers to think about more objectives, new alternatives and other possible states of the future
- Prompting for alternatives including extreme or unusual scenarios

The BetterBeliefs platform reduces biases in crowdsourcing in three ways, algorithmically, interactively and culturally—see Figure 12.

Devitt, S.K., Pearce, T.R., Chowdhury, A., Mengersen, K. (2021). A Bayesian social platform for inclusive and evidence-based decision making. [under peer review] M. Alfano, C. Klein and J de Ridder (Eds.). *Social Virtue Epistemology*. Routledge.

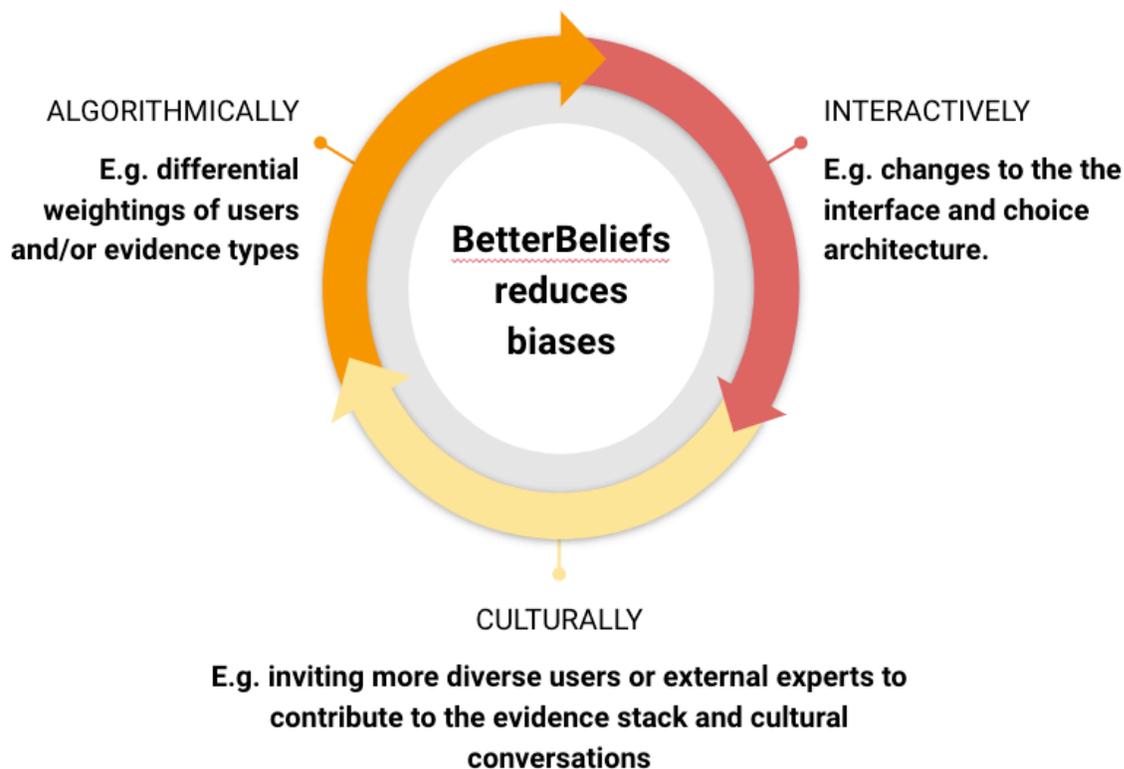

Figure 12 BetterBeliefs reduces biases algorithmically, interactively and culturally.

Changes to the user interface can reduce biases caused by the way information is displayed and choices are made. Biases can also be reduced culturally through the way the platform is used along with other workshop, ideation and research methods, training events and promotion of virtuous online behaviours by groups. Algorithmic methods to address bias include measurement of user interactions on the system and identifying biased or non-virtuous behaviours.

An example of algorithmic bias detection potential of the platform is using item-response methods (Embretson and Reise, 2013) to identify users that diverge from average response. In an analysis of one-use case of the platform, we could compare the success of ideas posted of skeptical users (those who tended to rate evidence as having less quality than the average user) with ideas posted by generous users (those who tended to rate evidence as having greater quality than average user). Some preliminary, correlative data (unfortunately unavailable in the public domain) suggests that a skeptical culture amongst groups who also engage in prolific hypothesis generation and evaluation may produce more successful ideas than more generous groups. But such a conjecture is purely speculative at this point and further experiments should be conducted to explore how diverse approaches to interaction on the platform affect the quality of outcomes for different purposes.

By encouraging virtuous epistemic behaviours (thinking of many ideas, justifying ideas with evidence and evaluating other people's ideas and evidence) and inhibiting unvirtuous behaviours, the platform ought to reduce a set of biases identified by Montibeller and von Winterfeldt (2015) including: anchoring bias, myopic problem representation, availability bias, omission of important variables, confirmation bias, and overconfidence bias—see Appendix 1. Biases reduced using the BetterBeliefs platform.

Increasing the number and diversity of hypotheses under consideration and encouraging individuals to justify them can improve decision making even if individual justifications are less than ideal (Oaksford et al., 2016). This comports with a Bayesian approach to evidence, which allows for evidence itself to vary in quality, so long as low-quality evidence is weighted less then higher quality evidence.

In addition to better hypotheses generation, there are significant benefits to decision makers of having a robust and dynamic set of evaluated hypotheses across teams and work hierarchies to amplify collective intelligence.

## Diversity of users

The norms of Bayesian epistemology recommends that more diverse stakeholders and more numerous independent evidential interactions on hypotheses will produce more defensible results to inform decision makers (Bovens & Hartmann, 2004; Devitt, 2013; Hajek & Hartmann, 2009). Diversity of stakeholders can be achieved in three different ways (Pinjani and Palvia, 2013):

1. demographic or surface-level diversity, e.g. age, sex, gender, race,
2. deep-level diversity, e.g. idiosyncratic attitudes, values, and preferences) or
3. functional diversity, non-overlapping knowledges and expertise in contributors, producing a larger knowledge base on which to draw.

Participants on a successful Bayesian social platform ought to encourage participation from all three kinds of diverse groups, as the likelihood of independence is increased by diversity. Not only did we seek functional diversity, but also to foster the ideas of those on the margins of groups and social networks. Weak ties between individuals have been shown to be good for innovation, where as strong ties between individuals have been shown to be good for productivity (Minor et al., 2017, Levin et al., 2011, Granovetter, 1973).

The platform supposes that the more competent, independent users on the platform considering ideas, the more likely a majority of those users are correct in accordance with Condorcet Jury Theorem (CJT). Condorcet Jury theorem supposes that incorporating the views of many minds (so long as they are competent and independent) will produce truthful propositions.

Not only is diversity important, but so is trust (Palvia, 2009). Contributors must trust that they are able to 'speak their mind' and given the benefit of the doubt, be treated with respect, be treated fairly, without unreasonable punitive actions being taken against them.

This method encourages an inclusive, yet evidence-based approach aiming for more reliable and useful results for stakeholders.

Devitt, S.K., Pearce, T.R., Chowdhury, A., Mengersen, K. (2021). A Bayesian social platform for inclusive and evidence-based decision making. [under peer review] M. Alfano, C. Klein and J de Ridder (Eds.). *Social Virtue Epistemology*. Routledge.

## Transparency and Access

Users and decision-makers can download data added to the platform including hypotheses, evidence items, degree of belief, weight of evidence, average quality of evidence, up votes, downvotes, vote count, rating count, total contributors and authors—see Figure 13. Users can choose real names or pseudonyms when they register. The privacy agreement on using the platform models best practise as per GDPR including making the privacy statement as clear as possible.

| | A | B | C | D | E | F | G | H | I | J | K | L |
|---|---|---|---|---|---|---|---|---|---|---|---|---|
| 1 | AddedOn | Title | Description | TagName | DegreeC | WeightOfl | AvgQua | UpVote | DownV | VoteCo | RatingCo | TotalC |
| 2 | 2018-10-09T | Nursing home elderley patient nutrition | 80% of elderly patients hospita | Nutrition | 0.9 | 7.6 | 3.8 | 5 | 0 | 5 | 4 | 6 |
| 3 | 2018-10-09T | Online ads are an effective way to chang | People do change behavior in r | Public Health | 0.8 | 7.6 | 3.7 | 2 | 0 | 2 | 3 | 3 |
| 4 | 2018-10-09T | We can improve human health & safety | What we eat is really importar | Environment | 0.8 | 8.5 | 4.0 | 9 | 1 | 10 | 4 | 11 |
| 5 | 2018-10-10T | Public health campaigns should target er | End-of-life care planning for pe | Aged Care | 0.8 | 6.8 | 3.5 | 4 | 0 | 4 | 2 | 5 |
| 6 | 2018-10-11T | Organizations should move to third gene | "In addition to general coping s | Mental Healt | 0.8 | 6.5 | 3.3 | 3 | 0 | 3 | 2 | 3 |
| 7 | 2018-10-11T | Investments are needed to prepare fami | Family members are often unp | Home Care | 0.8 | 8.3 | 4.2 | 5 | 1 | 6 | 3 | 6 |
| 8 | 2018-10-11T | The government should incentivise peopl | As people age, they may maint | Aged Care | 0.9 | 6.4 | 2.4 | 7 | 0 | 7 | 2 | 7 |
| 9 | 2018-10-11T | Increasing automation in hospital logistic | Can automation help achieve e | Hospital | 0.9 | 7.6 | 3.8 | 6 | 0 | 6 | 2 | 6 |
| 10 | 2018-10-11T | Reducing the stigma of psychological str | The stigma of psychological st | Mental Healt | 0.8 | 7.5 | 3.8 | 3 | 0 | 3 | 2 | 3 |
| 11 | | | | | | | | | | | | |

Figure 13. Sample of downloadable output from the BetterBeliefs platform (authors' names withheld)

## Identifying behaviours lacking virtue

The platform can use algorithmic means to identify online behaviours lacking value, such as:

> *Careless*: a user that endorse hypotheses or pieces of evidence without paying attention

> *Conformity:* a user being more likely to upvote hypothesis with high Degree of Belief (DoB) and give a high rank to those with high Weight of Evidence (WoE)

> *Authorship*: a user that downvotes or give low rank to refuting pieces of evidence on a hypothesis they entered and endorsed as well as the inclination to downvote or give low rank to pieces of evidence contrary hypotheses to author's

> *Group bias and manager fear bias*: Users that tend to favour an evidence/hypothesis from their area or added by their direct managers [or anyone higher in hierarchy].

> *Political coup*: a group of individuals acting cooperatively to achieve political ends. This may not be problematic if good and balanced evidence added. But, detecting such a bias could allow for early intervention on the coup

Once alerted to poor behaviours, moderators can intervene upon or remove users who are not confirming to community guidelines for online behaviours. There is still much work to be done to ensure moderators have appropriate checks on their own power to influence data production, manipulation and use. Being transparent about how data is generated and used to make decisions is critical in building and maintaining community trust. To date the BetterBeliefs platform has been used in organisational contexts where corporate, university or government ethics and decision-making is bound by explicit codes of conduct, human resource policy and legislative obligations.

Devitt, S.K., Pearce, T.R., Chowdhury, A., Mengersen, K. (2021). A Bayesian social platform for inclusive and evidence-based decision making. [under peer review] M. Alfano, C. Klein and J de Ridder (Eds.). *Social Virtue Epistemology*. Routledge.

# Discussion

Virtuous online digital communities seem like a great improvement over apathetic ones, so what could go wrong? In this section I outline some of the issues that are faced by online content providers and obligations they have to maintain a just and fair society as well as a knowledge-producing and truth-disseminating one. Key concerns include the tendency of platforms to exploit user attention and data to progress financial gain (particularly from advertising) to the detriment of user wellbeing; (Zuboff, 2019), the opaque use of surveillance and censorship (Lee and Scott-Baumann, 2020), lack of responsibility taken for damaging content posted to and disseminated on platforms in addition to a lack of regulatory oversight. We go through some of these issues in turn.

## Responsibility

Digital platforms have responsibility for both their function and their content. This means that they must have governance structures to evaluate and act on content shared on them if that content is misleading or false as well as causing harm or potentially causing harm. Facebook's Oversight Board is beginning to rule and have impacts on how Facebook manages content, such as the move to remove vaccine misinformation off the platform (Isaac, 2021). From responsibility also comes advocacy. Social platforms ought to take a stance on issues (such as public health) and justify behaviours based on this stance. We argue that supporting verifiable content and rejecting demonstrable falsehoods is a critical obligation of social platforms. However, content removal decisions ought to be scrutinised and held to a high standard, lest unwarranted censorship occurs.

## Free speech

Online platforms ought to encourage the free expression of ideas. Mark Zuckerberg has defended the value of free speech to justify *not* taking down posts with problematic content with the exception of posts that could lead to immediate direct physical harm to people on or off the platform. Free speech remains a controversial right as it is frequently misinterpreted as a freedom to say whatever an individual or group wishes to express. On the one hand, freedom is the founding value of the United States where many of the biggest social platforms arose, on the other hand, free speech is misunderstood as including falsehoods and asserting harmful propositions. The Oversight Board has called for Facebook to create more concrete policies that guide their content moderation decisions.

> The Board…found Facebook's misinformation and imminent harm rule… to be inappropriately vague and inconsistent with international human rights standards. A patchwork of policies found on different parts of Facebook's website make it difficult for users to understand what content is prohibited (Facebook Oversight Board, 2021).

Social platforms must abide by the legal obligations in the Sovereign nation within which they are based and abide by International legal frameworks that seek to minimise harms to others.

Devitt, S.K., Pearce, T.R., Chowdhury, A., Mengersen, K. (2021). A Bayesian social platform for inclusive and evidence-based decision making. [under peer review] M. Alfano, C. Klein and J de Ridder (Eds.). *Social Virtue Epistemology*. Routledge.

Freedom of expression ought to be endorsed in so far as it maintains authenticity, safety, privacy, dignity and the ability of others to also express themselves.

## Privacy

Online platforms ought to provide privacy to individuals and their content to the degree that users express a preference (Bernal, 2014). Such a view would defend a platform for allowing encryption to hide user content as well as allowing users to publicly promote their material. It would also obligate platforms not to conduct unnecessary surveillance or censorship upon users. Platforms must commit to security of data and information and to resolving data breaches quickly on behalf of users. There are ethical concerns with encryption, such as the wide dissemination with child pornography on communication apps that uses encryption. Material that might not be acceptable to the standards of society is likely to be shared via encrypted means. However, encryption also forms a necessary method and means by which citizens can mobilise against an unjust government or fight for their rights as citizens (Daly et al., 2019). Social platforms must remain vigilant with regards to best practice in privacy and security management and vow to continuously update their policies and action to meet the expectations of society and to progress a just and fair society.

## Data Rights and Data Activism

Social platforms ought to be GDPR compliant (or compliant with emerging local governance structures that promote user data rights) (European Parliament and Council, 2016). Data subjects ought to be able to request their data and to delete their data. Data activists ought to be able to access and make sense of social platform data creating new ways of knowing the world, creating data countercultures (Milan and Van der Velden, 2016). In general citizens ought to be more empowered to access and use data to progress their ends, particularly the most marginalised and disenfranchised (Daly et al., 2019).

Social platforms can learn from the emerging consensus in ethical AI with regards to how to consider the potential impacts of their technology on the society they serve—see Appendix 2. Comparison of AI Ethics Principles.

To date the BetterBeliefs platform has been used by organisations for closed groups for specific events including workshops, hackathons, design jams and stakeholder engagement for strategic policy setting. In closed settings, moderators and the platform designers have worked side-by-side to manage the ethics of platform use and disclosure to users. In the future, the platform team will need to carefully weigh up the excitement of expansion with the ethical risks such an expansion might reveal.

# Conclusion

Researchers have developed a technology that could be the first step in creating epistemic groups that use social platforms that are inclusive, responsive to evidence, limit punitive actions

Devitt, S.K., Pearce, T.R., Chowdhury, A., Mengersen, K. (2021). A Bayesian social platform for inclusive and evidence-based decision making. [under peer review] M. Alfano, C. Klein and J de Ridder (Eds.). *Social Virtue Epistemology*. Routledge.

and allow productive discord and respectful disagreement. BetterBeliefs improves evidence-based, collective ideation—a virtuous digital platform. Our design puts the hypothesis ahead of the document as the unit of information and evidence in the service of or arguing against hypotheses in accordance with the norms of Bayesian epistemology. The platform is designed to help reduce cognitive biases that emerge when groups produce too few hypotheses, hypotheses are too similar or conservative, collective knowledge is ignored, lost or under-utilised, evidence is not comprehensive or is drawn from conforming groups or contexts. Our platform encourages individuals to generate numerous and diverse hypotheses, prompts for different kinds of evidence to support or refute hypotheses, invites users to evaluate the quality of evidence, and scientifically calculates two kinds of metrics for the quality of hypotheses based on how people engage: a 'degree of belief' metric that measures how much confidence the group has in a hypothesis; and a 'weight of evidence' metric that measures how much evidence the group has considered for or against a hypothesis. The platform can be inclusive, intuitive and rewarding to use. However, while there is potential in using new types of social platforms, platform designs and providers must abide by emerging best practices in social platform governance and responsible innovation, ensuring responsibility, support of free speech, privacy by design, data rights and the opportunity for data activism.

APPENDIX 1 Biases reduced using the BetterBeliefs platform

**Anchoring bias** occurs when the estimation of a numerical value is based on an initial value (anchor), which is then insufficiently adjusted to provide the final answer.

Found in estimation tasks, pricing decisions and negotiations.

Ways to debias: avoiding anchors, providing multiple and counter anchors, and use experts with different anchors. Prompt employees to identify features of the target variable different than the anchor, or to consider reasons in conflict with the anchor.

**Availability bias** (or 'ease-of-recall') occurs when ease of recall dominates the assignment of probability to an event.

Found in frequency estimates, frequency of lethal events, and rare events that are anchored on recent examples.

Ways to debias include conducting probability training, provide counterexamples and provide statistics.

**Confirmation bias** occurs when there is a desire to confirm one's belief by selectively acquiring and using evidence.

Found in many settings such as information gathering, selection tasks, evidence updating and evaluation of one's own judgment. It has been shown in real-world contexts such as medical diagnostics, judicial reasoning and scientific thinking.

Ways to debias confirmation bias include using multiple experts with different points of view about hypotheses, challenging probability assessments with counterfactuals and probe evidence for alternative hypotheses.

**Myopic problem representation** occurs when an incomplete mental model creates an oversimplified problem representation.

Found when participants focus on a small number of alternatives, a small number of objectives, or a single future state of the world.

Ways to debias trying to encourage decision makers to think about more objectives, new alternatives and other possible states of the future.

**Omission of important variables** occurs when an important variable is overlooked.

Found in the definition of objectives, identification of decision alternatives and hypothesis generation. Ways to debias prompt for alternatives and objectives, ask for extreme or unusual scenarios or use group elicitation techniques.

Devitt, S.K., Pearce, T.R., Chowdhury, A., Mengersen, K. (2021). A Bayesian social platform for inclusive and evidence-based decision making. [under peer review] M. Alfano, C. Klein and J de Ridder (Eds.). *Social Virtue Epistemology*. Routledge.**Overconfidence bias** occurs when the decision makers provide estimates for a given parameter that are above the actual performance (overestimation) or when the range of variation they provide is too narrow (over precision).

Found frequently in quantitative estimates, such as in defence, legal, financial, and engineering decisions. Also present in judgments about the completeness of a hypothesis set.

Ways to debias include probability training; starting with extreme statistics (low and high), avoid central tendency anchors, use counterfactuals to challenge extremes. Use fixed value instead of fixed probability elicitations.

Devitt, S.K., Pearce, T.R., Chowdhury, A., Mengersen, K. (2021). A Bayesian social platform for inclusive and evidence-based decision making. [under peer review] M. Alfano, C. Klein and J de Ridder (Eds.). *Social Virtue Epistemology*. Routledge.

APPENDIX 2 Comparison of AI Ethics Principles

| Australian Government's AI Ethics Principles (Department of Industry Innovation and Science, 2019) | Principled Artificial Intelligence: A Map of Ethical and Rights Based Approaches | The global landscape of AI ethics guidelines (Jobin et al., 2019) |
|---|---|---|
| Human, social and environmental wellbeing: Throughout their lifecycle, AI systems should benefit individuals, society and the environment<br><br>Human-centred values: Throughout their lifecycle, AI systems should respect human rights, diversity, and the autonomy of individuals | Promotion of human values<br><br>Professional responsibility | Responsibility |
| Transparency and explainability: There should be transparency and responsible disclosure to ensure people know when they are being significantly impacted by an AI system, and can find out when an AI system is engaging with them | Human Control of Technology<br><br>Transparency | Transparency |
| Reliability and safety: Throughout their lifecycle, AI systems should reliably operate in accordance with their intended purpose | Fairness and non-discrimination<br><br>Safety and Security | Justice and fairness<br><br>Non-maleficence |

Devitt, S.K., Pearce, T.R., Chowdhury, A., Mengersen, K. (2021). A Bayesian social platform for inclusive and evidence-based decision making. [under peer review] M. Alfano, C. Klein and J de Ridder (Eds.). *Social Virtue Epistemology*. Routledge.

| | | |
|---|---|---|
| Fairness: Throughout their lifecycle, AI systems should be inclusive and accessible, and should not involve or result in unfair discrimination against individuals, communities or groups<br><br>Privacy protection and security: Throughout their lifecycle, AI systems should respect and uphold privacy rights and data protection, and ensure the security of data<br><br>Contestability: When an AI system significantly impacts a person, community, group or environment, there should be a timely process to allow people to challenge the use or output of the AI system | Privacy | Privacy |
| Accountability: Those responsible for the different phases of the AI system lifecycle should be identifiable and accountable for the outcomes of the AI systems, and human oversight of AI systems should be enabled | Accountability<br><br>Explainability | |